\documentclass{Interspeech}

\usepackage{amsmath,amssymb,amsfonts}
\usepackage{algorithmic}
\usepackage{graphicx}
\usepackage{textcomp}
\usepackage{xcolor}
\def\BibTeX{{\rm B\kern-.05em{\sc i\kern-.025em b}\kern-.08em
    T\kern-.1667em\lower.7ex\hbox{E}\kern-.125emX}}
\usepackage{soul,framed} 
\colorlet{shadecolor}{yellow}

\usepackage{array}
\usepackage{url}
\usepackage{cite}
\usepackage{booktabs}
\usepackage{balance} 
\usepackage{multirow}
\usepackage{multicol}
\usepackage{float}
\usepackage{bbold}
\usepackage{enumitem}
\usepackage{hyperref}
\usepackage{etoolbox,siunitx}
\robustify\bfseries
\usepackage{balance}

\usepackage{pifont}
\usepackage{scalerel}

\interspeechcameraready

\title{Plug-and-Play Co-Occurring Face Attention for Robust Audio-Visual \\Speaker Extraction}

\author[affiliation={1}]{Zexu}{Pan}
\author[affiliation={1}]{Shengkui}{Zhao}
\author[affiliation={2}]{Tingting}{Wang}
\author[affiliation={1}]{Kun}{Zhou}
\author[affiliation={1}]{Yukun}{Ma}
\author[affiliation={1}]{Chong}{Zhang}
\author[affiliation={1}]{Bin}{Ma}

\affiliation{Tongyi Lab}{Alibaba Group}{Singapore}
\affiliation{Nanjing University of Posts and Telecommunications}{NanJing}{China}
\email{zexu.pan@alibaba-inc.com}


\begin{document}

\maketitle

\begin{abstract}
Audio-visual speaker extraction isolates a target speaker's speech from a mixture speech signal conditioned on a visual cue, typically using the target speaker's face recording. However, in real-world scenarios, other co-occurring faces are often present on-screen, providing valuable speaker activity cues in the scene. In this work, we introduce a plug-and-play inter-speaker attention module to process these flexible numbers of co-occurring faces, allowing for more accurate speaker extraction in complex multi-person environments. We integrate our module into two prominent models: the AV-DPRNN and the state-of-the-art AV-TFGridNet. Extensive experiments on diverse datasets, including the highly overlapped VoxCeleb2 and sparsely overlapped MISP, demonstrate that our approach consistently outperforms baselines. Furthermore, cross-dataset evaluations on LRS2 and LRS3 confirm the robustness and generalizability of our method.
\end{abstract}

\section{Introduction}
In the real world, speech signals are often contaminated by overlapping background speech. This poses significant challenges for speech-based applications, including automatic speech recognition and speaker verification in human-computer interaction~\cite{Wang_aaai_2024}. Moreover, the presence of such noise degrades the quality of speech data, rendering its suitability for training large artificial intelligence models for speech understanding or generation tasks~\cite{QwenAudio}. Consequently, there is a critical need to extract clean speech from mixed signals, a challenge commonly known as the ``cocktail party problem"~\cite{cherry1953some}.

Speech separation (SS) is a widely studied approach to separate overlapping speech signals into individual tracks~\cite{hershey2016deep,wang2023tf,luo2019conv,zeghidour2020wavesplit,chen2017deep,von2022sasdr,pan2024paris}. However, SS often lacks speaker identity association. In addition, SS typically requires prior knowledge of the number of speakers for unified processing, which limits its capacity as the number of speakers increases. To address this, speaker extraction methods have been developed~\cite{vzmolikova2017learning,wang2019voicefilter,vzmolikova2019speakerbeam,sato2021multimodal}, inspired by human auditory attention mechanisms. These methods directly extract the target speaker's voice by conditioning on a reference signal, such as a reference speech signal~\cite{Chenglin2020spex,he2020speakerfilter}, visual recording~\cite{pan2020muse,wu2019time,pan2022seg,pan2023imaginenet}, or even brain activity~\cite{biss2020,pan2023neuroheed,pan2024neuroheed+}. Unlike speech separation, speaker extraction processes each speaker independently, avoiding identity and capacity limitations.

Among the auxiliary conditioning signals used in speaker extraction, visual cues have gained significant attention due to their robustness to acoustic noise. Typically, the synchronized face recordings extracted through face detection and tracking algorithms are used. Existing approaches often rely exclusively on the target speaker's visual cues, which the ambiguous visual-to-audio mappings (visemes to phonemes) frequently result in speaker confusion, over-suppression, and under-suppression errors in the extracted speech signals~\cite{wang2020voicefilter,zhao2022target,x-sepformer2023,pan2022hybrid}. In real-world scenarios such as meetings, interviews, or social gatherings, multiple faces are often present on-screen. These co-occurring faces provide valuable speech activity cues that can enhance the accuracy of target speaker extraction. For instance, if a co-occurring speaker is inactive, there is a higher likelihood that a speech signal belongs to the target speaker, and vice versa, particularly in sparsely overlapped multi-person interactions. By modeling cross-speaker information in such complex environments, these errors can be significantly reduced.

\begin{figure}[t]
\centering
\includegraphics[width=0.8\linewidth]{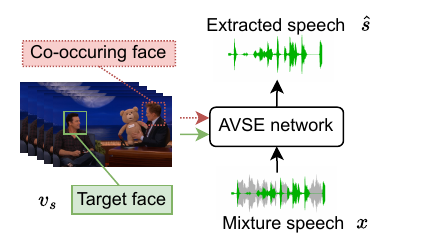}
\vspace{-4mm}
\caption{We explore the complementary speech activity cue offered by co-occurring on-screen face (red-dotted face box) when extracting the target (green non-dotted face box) speech in an audio-visual speaker extraction (AVSE) network.
}
\vspace{-6mm}
\label{fig:illustration}
\end{figure}

In this paper, we address this gap by proposing a lightweight, plug-and-play inter-speaker attention module designed to leverage the complementary speech activity cues provided by co-occurring faces. The attention mechanism is capable of handling a flexible number of on-screen faces. Our proposed module could be seamlessly integrated into various existing audio-visual speech extraction (AVSE) architectures and iteratively explores inter-speaker correlations within the mixed speech signals during the extraction process. 

To evaluate our approach, we integrate the proposed attention module into two representative AVSE models: (1) AV-DPRNN~\cite{usev21}, a popular and prominent dual-path time-domain masking-based model, and (2) AV-TFGridNet~\cite{pan2023avse}, a state-of-the-art framework that leverages time-frequency domain modeling for direct-signal estimation. We conduct extensive experiments on diverse datasets, including VoxCeleb2 (highly overlapped speech) and MISP (sparsely overlapped speech) for 2-speaker and 3-speaker scenarios, demonstrating that our method consistently outperforms baseline models with only 0.2 million additional parameters, regardless of the number of co-occurring faces available. Furthermore, cross-dataset evaluations on LRS2 and LRS3 confirm the robustness and generalizability of our approach, underscoring its practical applicability.

\section{Methodology}
\subsection{Fundamentals}
Denote $x$ as a multi-talker speech signal, which is the sum of the target speech signal $s$ and interfering speech signals $b_i$:
\begin{equation}
    x = s + \sum_{i=1}^{I}b_{i}
\end{equation}
where $i \in \{1,...,I\}$ denotes the index of interfering speakers, the audio-visual speaker extraction network extracts an estimate of $s$ denoted $\hat{s}$, from $x$, conditioned on the target speaker face recording $v_s$. Oftentimes, some interfering speakers' faces $v_{b_i}$ co-occur on the screen along with the target speaker, here we explore such complementary speech activity information. 

The majority of speaker extraction networks are inspired by speech separation architectures, typically comprised of a speech encoder, a speaker extractor, a speech decoder, and a visual encoder. The speaker extractor consists of multiple ($R$) repeated processing blocks, such as stacks of Temporal Convolutional Networks (TCNs) in Conv-TasNet~\cite{wu2019time,luo2019conv}, Dual-Path networks in DPRNN~\cite{luo2020dual,pan2022hybrid} or Sepformer~\cite{sepformer2021,av-speformer2023}, or frequency-domain attention in TF-GridNet~\cite{wang2023tf,pan2023avse}. In audio-visual speaker extraction, the output of the visual encoder is typically fused with one or all of these repeated blocks, enabling the network to leverage visual cues for improved speaker extraction performance.

\subsection{Plug-and-play inter-speaker attention module}
\begin{figure}[t]
\centering
\includegraphics[width=0.99\linewidth]{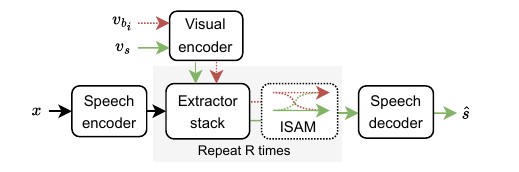}
\vspace{-3mm}
\caption{We introduce an optional (dotted box) inter-speaker attention module (ISAM) to compute attention to the co-occurring (dotted line) face activities. The plug-and-play ISAM is easily adaptable to the majority of AVSE networks.
}
\vspace{-2mm}
\label{fig:network}
\end{figure}

We introduce an Inter-Speaker Attention Module (ISAM) at the end of each repeated block in the speaker extractor, designed to be compatible with most speaker extraction networks. As illustrated in Figure~\ref{fig:network}, all available on-screen faces $v_s$ and $v_{b_i}$, are fed into the AVSE network (merged along the batch dimension), and the network is conditioned to extract the signal associated with each face independently through the visual encoder and extractor stacks. At the end of each extractor stack, we incorporate the ISAM, which enables the target speech embedding to query all speaker's embeddings, and vice versa. The ISAM consists of a single self-attention layer along the speaker axis, followed by a layer normalization. We employ self-attention rather than cross-attention because the embedding representations of co-occurring faces share the same distribution as the target face, as essentially the co-occurring faces could also be potential target faces in other words. The extractor stack and ISAM are repeated $R$ times to refine the extraction process.

To ensure the network remains robust in scenarios with no co-occurring faces, we occasionally bypass the ISAM during training. For scenarios involving an incomplete number of co-occurring faces, we employ random dropout of the co-occurring faces. This strategy enables the network to learn effective extraction even when co-occurring faces are absent or only partially available. Additionally, the ISAM is inherently robust to irrelevant on-screen faces (equivalent to silent faces in sparsely overlapping scenarios). This flexibility ensures that the module performs across a wide range of real-world conditions.

We integrate the ISAM into two popular AVSE models: AV-DPRNN~\cite{usev21} and AV-TFGridNet~\cite{pan2023avse}. AV-DPRNN, known for its balanced performance in speech extraction and computational complexity, has demonstrated strong results on both highly overlapping~\cite{pan2022hybrid} and sparsely overlapping speech~\cite{usev21}. In this model, we apply the ISAM to the embeddings after each stack of intra- and inter-chunk RNN processing. AV-TFGridNet, the current state-of-the-art AVSE model, operates in the frequency domain. Here, we apply the ISAM to every time-frequency bin across speakers after each stack of intra-frame, sub-band, and full-band processing. 
 
\subsection{Objective function}
We use the negative scale-invariant signal-to-noise ratio (SI-SNR)~\cite{le2019sdr} as the objective function for training:
\begin{equation}
    \label{eqa:loss_SI-SNR}
    \mathcal{L}_{\text{SI-SNR}}(s, \hat{s}) = - 20 \log_{10} \frac{|\frac{<\hat{s},s>}{|s|^2}s|}{|\hat{s} - \frac{<\hat{s},s>}{|s|^2}s\big|}
\end{equation}
The loss is applied to both the target speaker ($s$, $\hat{s}$), and the co-occuring speakers ($b_i$, $\hat{b_i}$).

\section{Experimental setup}

\subsection{Datasets}

We evaluate our proposed method and baselines using several audio-visual datasets: VoxCeleb2~\cite{chung2018voxceleb2}, LRS2~\cite{afouras2018deep}, LRS3~\cite{afouras2018lrs3}, and MISP~\cite{chen2022misp}. The VoxCeleb2 dataset is a multilingual, in-the-wild collection featuring diverse speakers. The LRS2 dataset comprises English speakers from BBC videos. The LRS3 dataset includes English speakers from TED videos, offering cleaner speech signals compared to VoxCeleb2 and LRS2. Lastly, the MISP dataset consists of Chinese speakers recorded in home conversation scenarios. These datasets collectively provide a comprehensive evaluation across varied languages, environments, and speech conditions.

The VoxCeleb2, LRS2, and LRS3 datasets consist exclusively of active speakers, making them suitable for simulating highly overlapped multi-talker speech. Following~\cite{pan2021reentry}, we mix the target speech with interference speech at a Signal-to-Noise Ratio (SNR) randomly set between $10$ dB and $-10$ dB. To maximize overlap, the longer speech clip is truncated to match the length of the shorter one. In contrast, the MISP dataset includes both active and inactive speakers, making it ideal for simulating sparsely overlapped multi-talker speech. Following~\cite{usev21}, we select non-overlapping single-speaker segments as sources for simulation, mixing the interference speech with the target speech at an SNR randomly set between $10$ dB and $-10$ dB. For this study, we simulate both 2-speaker and 3-speaker scenarios. All audio signals are sampled at $16$ kHz, while video signals are sampled at $25$ frames per second.

\subsection{Training}
For model implementation and experiments, we utilize the ClearerVoice-Studio toolkit~\footnote{\url{https://github.com/modelscope/ClearerVoice-Studio/}}. During training, we employ the Adam optimizer~\cite{kingma2015adam} and adopt a learning rate warmup strategy. Specifically, the learning rate (lr) is increased over the first $warmup_n = 15,000$ training steps according to the following formula:
\begin{equation}
\text{lr} = lr_{max} \cdot 0.001^{-1} \cdot 64^{-0.5} \cdot step_n \cdot warmup_n^{-1.5}
\end{equation}
where $step_n$ is the current step number. The maximum learning rate $lr_{max}$ is set to $0.001$ for the AV-DPRNN series of models and $0.0001$ for the AV-TF-GridNet series. After the warmup phase, the learning rate is halved if the validation loss does not improve for 6 consecutive epochs, and training stops if no improvement is observed within 10 subsequent epochs. All models are trained on four $80$ GB A800 GPUs using the NVIDIA platform, with an effective batch size of $24$ for the AV-DPRNN series and $16$ for the AV-TF-GridNet series.

\subsection{Network configuration}
The hyperparameter settings for AV-DPRNN follow~\cite{pan2022hybrid}, while those for AV-TFGridNet follow~\cite{pan2023avse}. For ISAM, the attention layer’s embedding size matches the hidden dimension of the respective backbone, with a feedforward dimension twice the embedding size. The number of attention heads is set to 1. AV-DPRNN has 15.3 million (M) parameters, increasing to 15.5M with ISAM, while AV-TFGridNet grows from 20.8M to 21.0M with ISAM
~\footnote{The model and training scripts are available at \url{https://github.com/modelscope/ClearerVoice-Studio}}. 

\begin{table}
    \centering
    \sisetup{
    detect-weight, 
    mode=text, 
    tight-spacing=true,
    round-mode=places,
    round-precision=1,
    table-format=2.1
    }
    \caption{Results on VoxCeleb2 2-speaker mixture dataset. v-spks denotes the number of face tracks observed by the model: 1-spk for the target speaker only, 2-spk for both speakers.}
    \vspace{-3mm}
    \addtolength{\tabcolsep}{-3pt}
    \resizebox{\linewidth}{!}{
    \begin{tabular}{ccc *{2}{S[round-precision=1,table-format=2.1]}*{2}{S[round-precision=2,table-format=1.2]}}
       \toprule
        Sys.    &Model                       &v-spks        &{SI-SNRi}  &{SNRi}     &{PESQi}    &STOIi\\
        \midrule
        1   &AV-DPRNN~\cite{usev21}  &1-spk         &11.5       &11.8   &0.88832    &0.23720\\
        \midrule
        2   &AV-DPRNN-SS             &2-spk         &10.69   &11.01   &0.771  &0.2231\\
        3   &AV-DPRNN-SE             &2-spk         &11.19  &11.505 &0.831  &0.2328\\
        \midrule
        \multirow{2}*{4}   &\multirow{2}*{AV-DPRNN-ISAM}      &1-spk         &11.7  &12.00  &0.9207 &0.2400\\
           &      &2-spk         &12.5   &12.8   &1.004  &0.2522\\
        \midrule
        5   &AV-DPRNN-ISAM$^{*}$                     &2-spk         &12.77  &13.07  &1.025  &0.2553\\
        \midrule
        \multirow{2}*{6}   &\multirow{2}*{AV-DPRNN-ISAM$^{\dagger}$}     &1-spk         &11.7   &12.02  &0.9048 &0.2395\\
           &     &2-spk         &12.36  &12.68  &0.964  &0.2488\\
        \midrule
         7  &AV-TFGridNet~\cite{pan2023avse}  &1-spk         &13.7448	&14.07	&1.454	&0.276\\
        \midrule
         \multirow{2}*{8}  &\multirow{2}*{AV-TFGridNet-ISAM}      &1-spk         &13.8622	&14.1803	&1.4813	&0.2785\\
           &      &2-spk         &14.4913	&14.7663	&1.566	&0.28842\\
        \bottomrule
    \end{tabular}
    }
    \addtolength{\tabcolsep}{3pt}
    \vspace{-2mm}
    \label{tab:vox_2mix}
\end{table}

\begin{table}
    \centering
    \sisetup{
    detect-weight, 
    mode=text, 
    tight-spacing=true,
    round-mode=places,
    round-precision=1,
    table-format=2.1
    }
    \caption{Results on VoxCeleb2 3-speaker mixture dataset. 1-spk means only the target speaker's face is visible, 2-spk includes the target and one random interfering speaker, and 3-spk includes all speakers.}
    \vspace{-3mm}
    \addtolength{\tabcolsep}{-3pt}
    \resizebox{\linewidth}{!}{
    \begin{tabular}{ccc *{2}{S[round-precision=1,table-format=2.1]}*{2}{S[round-precision=2,table-format=1.2]}}
       \toprule
        Sys.    &Model                       &v-spks        &{SI-SNRi}  &{SNRi}     &{PESQi}    &STOIi\\
        \midrule
        9   &AV-DPRNN~\cite{usev21}  &1-spk         &10.569    &11.085 &0.374239   &0.2910\\
        \midrule
        10   &AV-DPRNN-SE             &3-spk         &8.771  &9.2989 &0.26044    &0.2243\\
        \midrule
        \multirow{3}*{11}   &\multirow{3}*{AV-DPRNN-ISAM}      &1-spk         &11.4592 &12.00799   &0.45238    &0.3012\\
           &      &2-spk         &11.8942  &12.2681    &0.4614 &0.3089\\
           &      &3-spk         &13.311   &13.6896    &0.6074 &0.33737\\
       \midrule
         12  &AV-TFGridNet~\cite{pan2023avse}  &1-spk         &14.1558	&14.597	&0.892	&0.364    \\
        \midrule
         \multirow{3}*{13}  &\multirow{3}*{AV-TFGridNet-ISAM}      &1-spk         &14.2587	&14.7089	&0.928	&0.368\\
           &      &2-spk         &14.7764	&15.152	&0.9619	&0.376\\
           &      &3-spk         &15.5649	&15.937	&1.0807	&0.3897\\
        \bottomrule
    \end{tabular}
    }
    \addtolength{\tabcolsep}{3pt}
    \vspace*{-2mm}
    \label{tab:vox_3mix}
\end{table}

\begin{table}
    \centering
    \sisetup{
    detect-weight, 
    mode=text, 
    tight-spacing=true,
    round-mode=places,
    round-precision=1,
    table-format=2.1
    }
    \caption{Results on MISP 2-speaker mixture dataset.}
    \vspace{-3mm}
    \addtolength{\tabcolsep}{-3pt}
    \resizebox{\linewidth}{!}{
    \begin{tabular}{ccc *{2}{S[round-precision=1,table-format=2.1]}*{2}{S[round-precision=2,table-format=1.2]}}
       \toprule
        Sys.    &Model                       &v-spks        &{SI-SNRi}  &{SNRi}     &{PESQi}    &STOIi\\
        \midrule
         14  &AV-DPRNN~\cite{usev21}  &1-spk         &8.7417	&9.8583	&0.69035	&0.12003\\
        \midrule
         \multirow{2}*{15}  &\multirow{2}*{AV-DPRNN-ISAM}      &1-spk         &8.742	&9.9338	&0.6997	&0.1147\\
           &      &2-spk         &11.4122	&12.443	&0.8641	&0.1598\\
       \midrule
         16  &AV-TFGridNet~\cite{pan2023avse}  &1-spk         &10.3827	&11.6101	&1.0289	&0.1466\\
        \midrule
         \multirow{2}*{17}  &\multirow{2}*{AV-TFGridNet-ISAM}      &1-spk         &10.4683	&11.9547	&1.0777	&0.1474\\
           &      &2-spk         &13.3349	&14.4766	&1.25935	&0.18848\\
        \bottomrule
    \end{tabular}
    }
    \addtolength{\tabcolsep}{3pt}
    \vspace{-2mm}
    \label{tab:misp_2mix}
\end{table}

\begin{table}
    \centering
    \sisetup{
    detect-weight, 
    mode=text, 
    tight-spacing=true,
    round-mode=places,
    round-precision=1,
    table-format=2.1
    }
    \caption{Results on MISP 3-speaker mixture dataset.}
    \vspace{-3mm}
    \addtolength{\tabcolsep}{-3pt}
    \resizebox{\linewidth}{!}{
    \begin{tabular}{ccc *{2}{S[round-precision=1,table-format=2.1]}*{2}{S[round-precision=2,table-format=1.2]}}
       \toprule
        Sys.    &Model                       &v-spks        &{SI-SNRi}  &{SNRi}     &{PESQi}    &STOIi\\
        \midrule
         18  &AV-DPRNN~\cite{usev21}  &1-spk         &6.7032	&7.837	&0.2555	&0.1231\\
        \midrule
         \multirow{3}*{19}  &\multirow{3}*{AV-DPRNN-ISAM}      &1-spk         &7.8002	&9.0731	&0.36393	&0.1551\\
           &      &2-spk         &8.8677	&10.019	&0.4158	&0.1759\\
            &      &3-spk         &11.0363	&12.2172	&0.50586	&0.2175\\
       \midrule
         20  &AV-TFGridNet~\cite{pan2023avse}  &1-spk         &9.2911	&10.6014	&0.5829	&0.19429\\
        \midrule
         \multirow{3}*{21}  &\multirow{3}*{AV-TFGridNet-ISAM}      &1-spk         &9.2654	&10.6453	&0.59	&0.19752\\
           &      &2-spk         &9.7122	&11.0572	&0.62588	&0.2113\\
            &      &3-spk         &12.5354	&13.8709	&0.7663	&0.25732\\
        \bottomrule
    \end{tabular}
    }
    \addtolength{\tabcolsep}{3pt}
    \vspace{-2mm}
    \label{tab:misp_3mix}
\end{table}

\section{Result}
We evaluate the extracted speech signals using two sets of metrics: SI-SNRi and SNRi (in dB) for signal quality, and PESQi and STOIi for perceptual quality and intelligibility. All metrics are computed relative to the unprocessed multi-talker speech signal, the higher the better. We primarily compare SI-SNRi, as other metrics show similar trends. Each trained system is uniquely identified by a system number (Sys.) in all tables.

\subsection{Highly overlapped speech mixture}
We first evaluate ISAM using the highly overlapped VoxCeleb2 mixture dataset. Table~\ref{tab:vox_2mix} presents results on 2-speaker mixtures. The baseline System 1, a vanilla AV-DPRNN, achieves an SI-SNRi of $11.5$ dB. In contrast, System 2, which receives all faces as input and performs speech separation with two output streams, sees a drop in SI-SNRi to $10.7$ dB, likely due to the network's limited capacity to process excessive information. System 3, which also takes all faces as input but extracts only the target speaker, performs better at 11.2 dB SI-SNRi but still lags behind system 1.

With our proposed ISAM dropout training, System 4 achieves $11.7$ dB SI-SNRi when only the target speaker's face is available similarly to system 1, when an interfering face is also present (2-spk), SI-SNRi improves to $12.5$ dB, demonstrating ISAM’s effectiveness. As an upper bound, System 5 always employs ISAM and receives all speaker faces as input during training and inference, which achieves the highest $12.8$ dB SI-SNRi. System 6 is a variation of System 4, where instead of ISAM dropout, the target branch always passes through ISAM even when no co-occurring face is available. Its performance is similar to System 4, but we opt for System 4 due to its lower parameter count and reduced computational cost thanks to the ISAM dropout.

System 7 is AV-TFGridNet baseline, achieving an SI-SNRi of $13.7$ dB. System 8 is AV-TFGridNet with our ISAM. Due to the long training time of AV-TFGridNet models, all AV-TFGridNet-ISAM models in this paper are finetuned from the AV-TFGridNet baseline, whereas the AV-DPRNN-ISAM models are trained from scratch due to their simpler architecture and faster development cycle. For System 8, when only the target speaker is visible, the SI-SNRi is $13.9$ dB, but improves to $14.5$ dB when all speakers are observed.

Table~\ref{tab:vox_3mix} presents results on VoxCeleb2 3-speaker mixtures. The AV-DPRNN baseline (System 9) achieves an SI-SNRi of $10.6$ dB, while System 11, which incorporates ISAM, outperforms it in the 1-spk scenario with $11.5$ dB SI-SNRi. This improvement is likely due to the enhanced performance in the 2-spk and 3-spk scenarios, which also benefits the 1-spk case. When one interfering speaker is observed (2-spk), SI-SNRi improves to $11.9$ dB, and with all interference speakers present (3-spk), performance further increases to $13.3$ dB. For AV-TFGridNet, the baseline (System 12) achieves $14.2$ dB SI-SNRi. With ISAM, System 13 shows a steady improvement, increasing from $14.3$ dB (1-spk) to $14.8$ dB (2-spk) and reaching $15.6$ dB (3-spk) as more interfering speakers are observed.

\begin{table}
    \centering
    \sisetup{
    detect-weight, 
    mode=text, 
    tight-spacing=true,
    round-mode=places,
    round-precision=1,
    table-format=2.1
    }
    \caption{Results on LRS2 2-speaker mixture dataset.}
    \vspace{-3mm}
    \addtolength{\tabcolsep}{-3pt}
    \resizebox{\linewidth}{!}{
    \begin{tabular}{ccc *{2}{S[round-precision=1,table-format=2.1]}*{2}{S[round-precision=2,table-format=1.2]}}
       \toprule
        Sys.    &Model                       &v-spks        &{SI-SNRi}  &{SNRi}     &{PESQi}    &STOIi\\
        \midrule
         1  &AV-DPRNN~\cite{usev21}  &1-spk         &11.6842	&12.0464	&0.8096	&0.2355\\
        \midrule
         \multirow{2}*{4}  &\multirow{2}*{AV-DPRNN-ISAM}      &1-spk         &11.8616	&12.184	&0.846	&0.2395\\
           &      &2-spk         &12.821	&13.185	&0.9299	&0.25037\\
       \midrule
         7  &AV-TFGridNet~\cite{pan2023avse}  &1-spk         &14.367	&14.6911	&1.4402	&0.2773\\
        \midrule
         \multirow{2}*{8}  &\multirow{2}*{AV-TFGridNet-ISAM}      &1-spk         &14.3695	&14.7129	&1.4528	&0.2769\\
           &      &2-spk         &15.0343	&15.3628	&1.5522	&0.2854\\
        \bottomrule
    \end{tabular}
    }
    \addtolength{\tabcolsep}{3pt}
    \vspace{-2mm}
    \label{tab:lrs2_2mix}
\end{table}

\begin{table}
    \centering
    \sisetup{
    detect-weight, 
    mode=text, 
    tight-spacing=true,
    round-mode=places,
    round-precision=1,
    table-format=2.1
    }
    \caption{Results on LRS2 3-speaker mixture dataset.}
    \vspace{-3mm}
    \addtolength{\tabcolsep}{-3pt}
    \resizebox{\linewidth}{!}{
    \begin{tabular}{ccc *{2}{S[round-precision=1,table-format=2.1]}*{2}{S[round-precision=2,table-format=1.2]}}
       \toprule
        Sys.    &Model                       &v-spks        &{SI-SNRi}  &{SNRi}     &{PESQi}    &STOIi\\
        \midrule
         9  &AV-DPRNN~\cite{usev21}  &1-spk         &9.7881	&10.32	&0.29797	&0.261\\
        \midrule
         \multirow{3}*{11}  &\multirow{3}*{AV-DPRNN-ISAM}      &1-spk         &10.8649	&11.504	&0.3792	&0.2911\\
           &      &2-spk         &11.1758	&11.5744	&0.36542	&0.2951\\
            &      &3-spk         &13.1933	&13.64	&0.51567	&0.332\\
       \midrule
         12  &AV-TFGridNet~\cite{pan2023avse}  &1-spk         &13.9609	&14.4382	&0.7864	&0.3592\\
        \midrule
         \multirow{3}*{13}  &\multirow{3}*{AV-TFGridNet-ISAM}      &1-spk         &13.8892	&14.3859	&0.81058	&0.3594\\
           &      &2-spk         &14.4489	&14.833	&0.84767	&0.3675\\
            &      &3-spk         &15.7151	&16.1063	&1.0045	&0.3877\\
        \bottomrule
    \end{tabular}
    }
    \addtolength{\tabcolsep}{3pt}
    \vspace{-2mm}
    \label{tab:lrs2_3mix}
\end{table}

\subsection{Sparsely overlapped speech mixture}
Table~\ref{tab:misp_2mix} and Table~\ref{tab:misp_3mix} present results for 2-speaker and 3-speaker MISP mixtures, respectively. The results show a similar trend to those on the VoxCeleb2 mixture datasets, where ISAM consistently improves the performance of both AV-DPRNN and AV-TFGridNet across all conditions whenever any number of co-occurring faces are available.

We also observe that SI-SNRi on MISP is generally lower than on VoxCeleb2. This is because the sparsely overlapped mixtures pose a greater challenge: the network must jointly perform target speaker voice activity detection, and speaker separation when the target speaker is active. In contrast, in highly overlapped scenarios, the network only needs to focus on speaker separation, making the task comparatively easier. Nevertheless, ISAM provides even greater benefits in the sparsely overlapped scenario. For instance, on the 2-speaker mixtures, AV-TFGridNet-ISAM (System 8) shows a $0.6$ dB SI-SNRi improvement from 1-spk to 2-spk on the VoxCeleb2 mixture dataset in Table~\ref{tab:vox_2mix}, while AV-TFGridNet-ISAM (System 17) achieves a $2.8$ dB  SI-SNRi improvement on the MISP mixture dataset in Table~\ref{tab:misp_2mix}. Similarly, on the 3-speaker mixtures, AV-TFGridNet-ISAM (System 13) improves  SI-SNRi by $1.3$ dB from 1-spk to 3-spk on VoxCeleb2 mixture dataset in Table~\ref{tab:vox_3mix}, whereas AV-TFGridNet-ISAM (System 21) achieves a $3.2$ dB SI-SNRi improvement on MISP mixture dataset in Table~\ref{tab:misp_3mix}. These results highlight the effectiveness of ISAM in handling sparsely overlapped mixtures, where leveraging co-occurring faces provides even greater advantages.

\begin{table}
    \centering
    \sisetup{
    detect-weight, 
    mode=text, 
    tight-spacing=true,
    round-mode=places,
    round-precision=1,
    table-format=2.1
    }
    \caption{Results on LRS3 2-speaker mixture dataset.}
    \vspace{-3mm}
    \addtolength{\tabcolsep}{-3pt}
    \resizebox{\linewidth}{!}{
    \begin{tabular}{ccc *{2}{S[round-precision=1,table-format=2.1]}*{2}{S[round-precision=2,table-format=1.2]}}
       \toprule
        Sys.    &Model                       &v-spks        &{SI-SNRi}  &{SNRi}     &{PESQi}    &STOIi\\
        \midrule
         1  &AV-DPRNN~\cite{usev21}  &1-spk         &13.1027	&13.4045	&1.00104	&0.236\\
        \midrule
         \multirow{2}*{4}  &\multirow{2}*{AV-DPRNN-ISAM}      &1-spk         &13.2824	&13.5499	&1.0404	&0.23973\\
           &      &2-spk         &14.2243	&14.512	&1.1471	&0.2493\\
       \midrule
         7  &AV-TFGridNet~\cite{pan2023avse}  &1-spk         &16.2282	&16.4658	&1.73288	&0.27311\\
        \midrule
         \multirow{2}*{8}  &\multirow{2}*{AV-TFGridNet-ISAM}      &1-spk         &16.244	&16.4874	&1.7686	&0.2731\\
           &      &2-spk         &16.9063	&17.1142	&1.8624	&0.27968\\
        \bottomrule
    \end{tabular}
    }
    \addtolength{\tabcolsep}{3pt}
    \vspace{-2mm}
    \label{tab:lrs3_2mix}
\end{table}

\begin{table}
    \centering
    \sisetup{
    detect-weight, 
    mode=text, 
    tight-spacing=true,
    round-mode=places,
    round-precision=1,
    table-format=2.1
    }
    \caption{Results on LRS3 3-speaker mixture dataset.}
    \vspace{-3mm}
    \addtolength{\tabcolsep}{-3pt}
    \resizebox{\linewidth}{!}{
    \begin{tabular}{ccc *{2}{S[round-precision=1,table-format=2.1]}*{2}{S[round-precision=2,table-format=1.2]}}
       \toprule
        Sys.    &Model                       &v-spks        &{SI-SNRi}  &{SNRi}     &{PESQi}    &STOIi\\
        \midrule
         9  &AV-DPRNN~\cite{usev21}  &1-spk         &11	&11.413	&0.3997	&0.2719\\
        \midrule
         \multirow{3}*{11}  &\multirow{3}*{AV-DPRNN-ISAM}      &1-spk         &12.0574	&12.542	&0.5048	&0.2963\\
           &      &2-spk         &12.5208	&12.8	&0.5059	&0.3041\\
            &      &3-spk         &14.6753	&14.984	&0.70425	&0.33851\\
       \midrule
         12  &AV-TFGridNet~\cite{pan2023avse}  &1-spk         &15.2433	&15.606	&0.9877	&0.3564\\
        \midrule
         \multirow{3}*{13}  &\multirow{3}*{AV-TFGridNet-ISAM}      &1-spk         &15.1143	&15.4765	&1.0222	&0.3547\\
           &      &2-spk         &15.8752	&16.1287	&1.07827	&0.36668\\
            &      &3-spk         &17.222	&17.44867	&1.2627	&0.38352\\
        \bottomrule
    \end{tabular}
    }
    \addtolength{\tabcolsep}{3pt}
    \vspace{-2mm}
    \label{tab:lrs3_3mix}
\end{table}

\subsection{Cross-dataset evaluation}
We also perform cross-dataset evaluations to assess the generalizability of ISAM. Table~\ref{tab:lrs2_2mix} and Table~\ref{tab:lrs2_3mix} present results for models trained on the VoxCeleb2 mixture but tested on the LRS2 mixture dataset. LRS2 is similar to VoxCeleb2, except that it contains only English speech, while VoxCeleb2 is multilingual. The results show that performance on LRS2 closely aligns with that on VoxCeleb2, with ISAM achieving similar performance gains over the baseline.

Additionally, Table~\ref{tab:lrs3_2mix} and Table~\ref{tab:lrs3_3mix} present results for models trained on VoxCeleb2 but tested on LRS3. Since LRS3 is a cleaner dataset than VoxCeleb2, the overall performance is higher. The performance gains of ISAM follow a similar trend to those observed on VoxCeleb2, further demonstrating its effectiveness across different datasets.

Across all datasets: VoxCeleb2, MISP, LRS2, and LRS3, we consistently observe that for 3-speaker mixtures, the performance gain from 1-spk to 2-spk is smaller than from 2-spk to 3-spk. This suggests that observing all faces provides the maximum performance improvement. Nevertheless, the 1-spk to 2-spk improvement remains significant, demonstrating that our model is robust to varying numbers of observed faces and effectively leverages available visual information.

\section{Conclusion}
In conclusion, we introduce the Inter-Speaker Attention Module (ISAM), a novel approach that enables audio-visual speaker extraction to utilize complementary speech activity cues from a flexible number of co-occurring on-screen faces. The ISAM not only enhances performance on highly overlapped speech mixtures but also achieves significant improvements on more challenging sparsely overlapped scenarios. Furthermore, our method demonstrates strong robustness and generalizability in cross-dataset evaluations, underscoring its effectiveness in real-world applications.


\bibliographystyle{IEEEtran}
\bibliography{mybib}

\begin{thebibliography}{10}
\providecommand{\url}[1]{#1}
\csname url@samestyle\endcsname
\providecommand{\newblock}{\relax}
\providecommand{\bibinfo}[2]{#2}
\providecommand{\BIBentrySTDinterwordspacing}{\spaceskip=0pt\relax}
\providecommand{\BIBentryALTinterwordstretchfactor}{4}
\providecommand{\BIBentryALTinterwordspacing}{\spaceskip=\fontdimen2\font plus
\BIBentryALTinterwordstretchfactor\fontdimen3\font minus \fontdimen4\font\relax}
\providecommand{\BIBforeignlanguage}[2]{{%
\expandafter\ifx\csname l@#1\endcsname\relax
\typeout{** WARNING: IEEEtran.bst: No hyphenation pattern has been}%
\typeout{** loaded for the language `#1'. Using the pattern for}%
\typeout{** the default language instead.}%
\else
\language=\csname l@#1\endcsname
\fi
#2}}
\providecommand{\BIBdecl}{\relax}
\BIBdecl

\bibitem{Wang_aaai_2024}
J.~Wang, Z.~Pan, M.~Zhang, R.~T. Tan, and H.~Li, ``Restoring speaking lips from occlusion for audio-visual speech recognition,'' in \emph{Proc. AAAI}, vol.~38, 2024.

\bibitem{QwenAudio}
Y.~Chu, J.~Xu, X.~Zhou, Q.~Yang, S.~Zhang, Z.~Yan, C.~Zhou, and J.~Zhou, ``{Qwen-Audio}: Advancing universal audio understanding via unified large-scale audio-language models,'' \emph{arXiv preprint arXiv:2311.07919}, 2023.

\bibitem{cherry1953some}
E.~C. Cherry, ``Some experiments on the recognition of speech, with one and with two ears,'' \emph{The Journal of the acoustical society of America}, vol.~25, no.~5, pp. 975--979, 1953.

\bibitem{hershey2016deep}
J.~R. {Hershey}, Z.~{Chen}, J.~{Le Roux}, and S.~{Watanabe}, ``Deep clustering: Discriminative embeddings for segmentation and separation,'' in \emph{Proc. ICASSP}, 2016, pp. 31--35.

\bibitem{wang2023tf}
Z.-Q. Wang, S.~Cornell, S.~Choi, Y.~Lee, B.-Y. Kim, and S.~Watanabe, ``{TF-GridNet}: Making time-frequency domain models great again for monaural speaker separation,'' in \emph{Proc. ICASSP}, 2023.

\bibitem{luo2019conv}
Y.~{Luo} and N.~{Mesgarani}, ``Conv-{TasNet}: Surpassing ideal time–frequency magnitude masking for speech separation,'' \emph{IEEE/ACM Trans. Audio, Speech, Lang. Process.}, vol.~27, no.~8, pp. 1256--1266, 2019.

\bibitem{zeghidour2020wavesplit}
N.~Zeghidour and D.~Grangier, ``Wavesplit: End-to-end speech separation by speaker clustering,'' \emph{IEEE/ACM Trans. Audio, Speech, Lang. Process.}, vol.~29, pp. 2840--2849, 2021.

\bibitem{chen2017deep}
Z.~Chen, Y.~Luo, and N.~Mesgarani, ``Deep attractor network for single-microphone speaker separation,'' in \emph{Proc. ICASSP}, 2017, pp. 246--250.

\bibitem{von2022sasdr}
T.~von Neumann, K.~Kinoshita, C.~Boeddeker, M.~Delcroix, and R.~Haeb-Umbach, ``{SA-SDR}: A novel loss function for separation of meeting style data,'' in \emph{Proc. ICASSP}, 2022, pp. 6022--6026.

\bibitem{pan2024paris}
Z.~Pan, G.~Wichern, F.~G. Germain, K.~Saijo, and J.~Le~Roux, ``{PARIS}: {Pseudo-AutoRegressIve} siamese training for online speech separation,'' in \emph{Proc. Interspeech}, 2024.

\bibitem{vzmolikova2017learning}
K.~{Žmolíková}, M.~{Delcroix}, K.~{Kinoshita}, T.~{Higuchi}, A.~{Ogawa}, and T.~{Nakatani}, ``Learning speaker representation for neural network based multichannel speaker extraction,'' in \emph{Proc. ASRU}, 2017, pp. 8--15.

\bibitem{wang2019voicefilter}
Q.~Wang, H.~Muckenhirn, K.~Wilson, P.~Sridhar, Z.~Wu, J.~R. Hershey, R.~A. Saurous, R.~J. Weiss, Y.~Jia, and I.~L. Moreno, ``{VoiceFilter}: Targeted voice separation by speaker-conditioned spectrogram masking,'' in \emph{Proc. Interspeech}, 2019, pp. 2728--2732.

\bibitem{vzmolikova2019speakerbeam}
K.~{Žmolíková}, M.~{Delcroix}, K.~{Kinoshita}, T.~{Ochiai}, T.~{Nakatani}, L.~{Burget}, and J.~{Černocký}, ``Speaker{B}eam: Speaker aware neural network for target speaker extraction in speech mixtures,'' \emph{IEEE Journal of Selected Topics in Signal Processing}, vol.~13, no.~4, pp. 800--814, 2019.

\bibitem{sato2021multimodal}
H.~Sato, T.~Ochiai, K.~Kinoshita, M.~Delcroix, T.~Nakatani, and S.~Araki, ``Multimodal attention fusion for target speaker extraction,'' in \emph{Proc. SLT}, 2021, pp. 778--784.

\bibitem{Chenglin2020spex}
C.~{Xu}, W.~{Rao}, E.~S. {Chng}, and H.~{Li}, ``Sp{E}x: Multi-scale time domain speaker extraction network,'' \emph{IEEE/ACM Trans. Audio, Speech, Lang. Process.}, vol.~28, pp. 1370--1384, 2020.

\bibitem{he2020speakerfilter}
S.~{He}, H.~{Li}, and X.~{Zhang}, ``{Speakerfilter}: Deep learning-based target speaker extraction using anchor speech,'' in \emph{Proc. ICASSP}, 2020, pp. 376--380.

\bibitem{pan2020muse}
Z.~Pan, R.~Tao, C.~Xu, and H.~Li, ``{MuSE}: Multi-modal target speaker extraction with visual cues,'' in \emph{Proc. ICASSP}, 2021, pp. 6678--6682.

\bibitem{wu2019time}
J.~{Wu}, Y.~{Xu}, S.~{Zhang}, L.~{Chen}, M.~{Yu}, L.~{Xie}, and D.~{Yu}, ``Time domain audio visual speech separation,'' in \emph{Proc. ASRU}, 2019, pp. 667--673.

\bibitem{pan2022seg}
Z.~Pan, X.~Qian, and H.~Li, ``Speaker extraction with co-speech gestures cue,'' \emph{IEEE Signal Process. Lett.}, vol.~29, pp. 1467--1471, 2022.

\bibitem{pan2023imaginenet}
Z.~Pan, W.~Wang, M.~Borsdorf, and H.~Li, ``{ImagineNet}: Target speaker extraction with intermittent visual cue through embedding inpainting,'' in \emph{Proc. ICASSP}, 2023.

\bibitem{biss2020}
E.~Ceolini, J.~Hjortkj{\ae}r, D.~D. Wong, J.~O’Sullivan, V.~S. Raghavan, J.~Herrero, A.~D. Mehta, S.-C. Liu, and N.~Mesgarani, ``Brain-informed speech separation for enhancement of target speaker in multitalker speech perception,'' \emph{NeuroImage}, vol. 223, p. 117282, 2020.

\bibitem{pan2023neuroheed}
Z.~Pan, M.~Borsdorf, S.~Cai, T.~Schultz, and H.~Li, ``Neuro{H}eed: Neuro-steered speaker extraction using {EEG} signals,'' \emph{IEEE/ACM Trans. Audio, Speech, Lang. Process.}, 2024.

\bibitem{pan2024neuroheed+}
Z.~Pan, G.~Wichern, F.~G. Germain, S.~Khurana, and J.~Le~Roux, ``{NeuroHeed+}: Improving neuro-steered speaker extraction with joint auditory attention detection,'' in \emph{Proc. ICASSP}, 2024, pp. 11\,456--11\,460.

\bibitem{wang2020voicefilter}
Q.~Wang, I.~L. Moreno, M.~Saglam, K.~Wilson, A.~Chiao, R.~Liu, Y.~He, W.~Li, J.~Pelecanos, M.~Nika \emph{et~al.}, ``{VoiceFilter-Lite}: Streaming targeted voice separation for on-device speech recognition,'' \emph{Proc. Interspeech}, pp. 2677--2681, 2020.

\bibitem{zhao2022target}
Z.~Zhao, D.~Yang, R.~Gu, H.~Zhang, and Y.~Zou, ``Target confusion in end-to-end speaker extraction: Analysis and approaches,'' \emph{Proc. Interspeech}, 2022.

\bibitem{x-sepformer2023}
K.~Liu, Z.~Du, X.~Wan, and H.~Zhou, ``{X-SepFormer}: End-to-end speaker extraction network with explicit optimization on speaker confusion,'' in \emph{Proc. ICASSP}, 2023, pp. 1--5.

\bibitem{pan2022hybrid}
Z.~Pan, M.~Ge, and H.~Li, ``A hybrid continuity loss to reduce over-suppression for time-domain target speaker extraction,'' in \emph{Proc. Interspeech}, 2022, pp. 1786--1790.

\bibitem{usev21}
------, ``{USEV}: Universal speaker extraction with visual cue,'' \emph{IEEE/ACM Trans. Audio, Speech, Lang. Process.}, vol.~30, pp. 3032--3045, 2022.

\bibitem{pan2023avse}
Z.~Pan, G.~Wichern, Y.~Masuyama, F.~G. Germain, S.~Khurana, C.~Hori, and J.~Le~Roux, ``Scenario-aware audio-visual {TF-Gridnet} for target speech extraction,'' in \emph{Proc. ASRU}, 2023.

\bibitem{luo2020dual}
Y.~{Luo}, Z.~{Chen}, and T.~{Yoshioka}, ``{Dual-Path RNN}: Efficient long sequence modeling for time-domain single-channel speech separation,'' in \emph{Proc. ICASSP}, 2020, pp. 46--50.

\bibitem{sepformer2021}
C.~Subakan, M.~Ravanelli, S.~Cornell, M.~Bronzi, and J.~Zhong, ``Attention is all you need in speech separation,'' in \emph{Proc. ICASSP}, 2021, pp. 21--25.

\bibitem{av-speformer2023}
J.~Lin, X.~Cai, H.~Dinkel, J.~Chen, Z.~Yan, Y.~Wang, J.~Zhang, Z.~Wu, Y.~Wang, and H.~Meng, ``[av-sepformer]: Cross-attention sepformer for audio-visual target speaker extraction,'' in \emph{Proc. ICASSP}, 2023, pp. 1--5.

\bibitem{le2019sdr}
J.~Le~Roux, S.~Wisdom, H.~Erdogan, and J.~R. Hershey, ``{SDR}--half-baked or well done?'' in \emph{Proc. ICASSP}, 2019, pp. 626--630.

\bibitem{chung2018voxceleb2}
J.~S. Chung, A.~Nagrani, and A.~Zisserman, ``Vox{C}eleb2: Deep speaker recognition,'' \emph{Proc. Interspeech}, pp. 1086--1090, 2018.

\bibitem{afouras2018deep}
T.~Afouras, J.~S. Chung, A.~Senior, O.~Vinyals, and A.~Zisserman, ``Deep audio-visual speech recognition,'' \emph{IEEE Trans. Pattern Anal. Mach. Intell.,}, 2018.

\bibitem{afouras2018lrs3}
T.~Afouras, J.~S. Chung, and A.~Zisserman, ``{LRS3-TED}: a large-scale dataset for visual speech recognition,'' \emph{arXiv preprint arXiv:1809.00496}, 2018.

\bibitem{chen2022misp}
H.~Chen, H.~Zhou, J.~Du, C.-H. Lee, J.~Chen, S.~Watanabe, S.~M. Siniscalchi, O.~Scharenborg, D.-Y. Liu, B.-C. Yin, J.~Pan, J.-Q. Gao, and C.~Liu, ``The first multimodal information based speech processing ({MISP}) challenge: Data, tasks, baselines and results,'' in \emph{Proc. ICASSP}, 2022.

\bibitem{pan2021reentry}
Z.~Pan, R.~Tao, C.~Xu, and H.~Li, ``Selective listening by synchronizing speech with lips,'' \emph{IEEE/ACM Trans. Audio, Speech, Lang. Process.}, vol.~30, pp. 1650--1664, 2022.

\bibitem{kingma2015adam}
D.~P. Kingma and J.~Ba, ``Adam, a method for stochastic optimization,'' in \emph{Proc. ICLR}, vol. 1412, 2015.

\end{thebibliography}

\end{document}